\documentclass[12pt]{spieman}  
\usepackage{amsmath,amsfonts,amssymb}
\usepackage{graphicx}
\usepackage{setspace}
\usepackage{tocloft}
\usepackage{lineno}
\usepackage{xcolor}

\title{Mapping Magnetic Fields from Clouds to Cores with PRIMAger}

\author[a,*]{Kate Pattle}
\author[a]{Janik Karoly}
\author[a]{Lorna Buhil Findlay}
\author[b,c]{Simon Coud\'e}
\author[d]{Brandon S. Hensley}
\author[e,f]{Paulo C. Cortes}
\author[g,h]{James Di Francesco}
\author[i,j]{Valentin J. M. Le Gouellec}
\author[k,l]{Enrique Lopez-Rodriguez}
\author[m]{Fabien Louvet}
\affil[a]{Department of Physics and Astronomy, University College London, Gower Street, London WC1E 6BT, UK}
\affil[b]{Department of Earth, Environment, and Physics, Worcester State University, Worcester, MA 01602, USA}
\affil[c]{Harvard-Smithsonian Center for Astrophysics, 60 Garden Street, Cambridge, MA, 02138, USA}
\affil[d]{Jet Propulsion Laboratory, California Institute of Technology, 4800 Oak Grove Drive, Pasadena, CA 91109, USA}
\affil[e]{Joint ALMA Observatory, Alonso de C\'ordova 3107, Vitacura, Santiago, Chile}
\affil[f]{National Radio Astronomy Observatory, 520 Edgemont Road, Charlottesville, VA 22903, USA}
\affil[g]{Herzberg Astronomy and Astrophysics, National Research Council of Canada, 5071 West Saanich Road, Victoria, BC, V9E 2E7, Canada}
\affil[h]{Department of Physics and Astronomy, University of Victoria, 3800 Finnerty Road, Elliott Building, Room 101, Victoria, BC, V8P 5C2, Canada}
\affil[i]{Institut de Ci\'encies de l'Espai (ICE-CSIC), Campus UAB, Can Magrans S/N, E-08193 Cerdanyola del Vall\`es, Catalonia, Spain}
\affil[j]{Institut d'Estudis Espacials de Catalunya (IEEC), c/Gran Capita, 2-4, E-08034 Barcelona, Catalonia, Spain}
\affil[k]{Kavli Institute for Particle Astrophysics \& Cosmology (KIPAC), Stanford University, Stanford, CA 94305, USA}
\affil[l]{Department of Physics \& Astronomy, University of South Carolina, Columbia, SC 29208, USA}
\affil[m]{Univ. Grenoble Alpes, CNRS, IPAG, 38000 Grenoble, France}

\newcommand{\mnras}{MNRAS}
\newcommand{\aap}{A\&A}
\newcommand{\nat}{Nature}

\newcommand{\apj}{ApJ}
\newcommand{\apjs}{ApJS}
\newcommand{\apjl}{ApJL}
\newcommand{\aj}{AJ}

\newcommand{\araa}{ARA\&A}

\newcommand{\nar}{New Astronomy Reviews}
\newcommand{\prl}{Phys. Rev. Letters}

\cftpagenumbersoff{figure}
\cftpagenumbersoff{table} 
\begin{document} 
\maketitle

\begin{abstract}
High-resolution, wide-area mapping of magnetic field geometries within molecular clouds, and the star-forming filaments and cores within them, is crucial in order to understand the role of magnetic fields in the star formation process.  We therefore propose an unbiased survey of star-forming molecular clouds within 0.5 kpc of the Earth in polarized light with the PRIMAger Polarimetry Imager.  We will map magnetic fields over entire molecular clouds at linear resolutions of $\sim10^{-3}-10^{-2}$ pc ($\sim10^{3}-10^{4}$ au) in PRIMAger Bands PPI1 -- PPI4, thereby resolving magnetic field structure both within individual star-forming filaments and cores, and in the most diffuse regions of molecular clouds.  These multi-wavelength polarimetric observations will allow us to systematically investigate both the wide range of open questions about the role of magnetic fields in star formation and the evolution of the interstellar medium, and interstellar dust grain properties.  The time required to map the area observed by the \textit{Herschel} Gould Belt Survey (160 deg$^{2}$) to the cirrus confusion limit in polarized light is 170 hours.  This will give a 5-$\sigma$ detection of 20\% polarized low-density cirrus emission, with surface brightnesses in polarized intensity of 1.0--2.4\,MJy/sr across the PRIMAger bands, and will ensure detection of polarized emission at all higher column densities.  This time estimate can be simply scaled up in order to map magnetic fields in a larger sample of molecular clouds, including more distant regions of higher-mass star formation.

\end{abstract}


{\noindent \footnotesize\textbf{*}Kate Pattle, \linkable{k.pattle@ucl.ac.uk} }


\section{Introduction}
\label{sect:intro}  

Star formation is one of the most important unsolved problems in modern astrophysics, because how and where stars form has consequences across the discipline, from galaxy evolution to planet formation.  A key unknown in studies of star formation and the physics of the interstellar medium (ISM) is that of the role and relative importance of magnetic fields\cite{pattle2023}.

Stars form from the gravitational collapse of overdensities within clouds of molecular hydrogen gas (hereafter `molecular clouds'), the densest and coldest phase of the interstellar medium.  These molecular clouds are threaded on all size scales by ordered magnetic fields, which on the largest scales are inherited from the Milky Way's galactic-dynamo-driven magnetic field\cite{beck1996}, and on the smallest scales provide the initial conditions for magnetised accretion and jet launching in protostellar discs \cite{tsukamoto2023}.  High-resolution, wide-area mapping of the magnetic field geometries within these clouds, filaments and cores is therefore crucial in order to understand the dynamic role that these fields play in the star formation process.

The energetic importance of magnetic fields in molecular clouds is not well-constrained, and seems to vary with size scale and environment\cite{pattle2023}.  Magnetic fields appear to transition from dominating over gravity and being near equipartition with turbulent gas motions on large scales in low-density cloud envelopes\cite{soler2019} on size scales $\sim 10$\,pc or larger, to being energetically sub-dominant -- albeit perhaps only mildly so -- within dense, gravitationally unstable star-forming cores\cite{myers2021,ching2022} with size scales $\lesssim0.1$\,pc.  They may play an important role in the formation of cloud substructure, and particularly in the formation of filamentary structure\cite{palmeirim2013}, and may direct the accretion of material along filaments onto high-mass star-forming hubs\cite{pillai2020}.

Many key questions about the role and relative importance of magnetic fields in the star formation process remain to be answered, including, but not limited to: 
\begin{enumerate}
    \item The role of magnetic fields in structure formation and fragmentation within molecular clouds.
    \item Whether molecular clouds undergo a transition from sub- to super-criticality (from magnetically dominated to gravitationally collapsing), and if so, whether there is a characteristic size or density scale on which this occurs. 
    \item The existence, or otherwise, of distinct weak- and strong-field modes of star formation, and if such distinct modes do exist, whether or not they produce differences in the measurable outcomes of the star formation process, such as star formation efficiency or the Initial Mass Function.
    \item How magnetic fields and protostellar and stellar feedback interact, and what role this interaction plays in regulating star formation.
\end{enumerate}
Systematic investigation of these questions requires an unbiased survey of magnetic field morphologies in star-forming molecular clouds with an instrument capable both of mapping entire clouds and of resolving individual star-forming filaments and cores.  For this latter requirement, angular resolutions of tens of arcseconds are required: the Jeans length is $\sim 10^{-2}$\,pc for 10$^{6}$\,cm$^{-3}$ gas at a temperature of 15\,K, or $\sim 40^{\prime\prime}$ at a distance of 500\,pc.

As we demonstrate in this paper, the Probe far-Infrared Mission for Astrophysics (PRIMA) will give us an unprecedented opportunity to accomplish the goal described above by using the PRIMAger instrument's Polarimetry Imager to map nearby molecular clouds in polarized light.  Magnetic fields in star-forming clouds can be traced by observing linearly polarized continuum emission from the interstellar dust grains which are mixed with the molecular gas of the clouds.  These dust grains preferentially align with their minor axes perpendicular to the local magnetic field direction\cite{andersson2015}, and so their thermal emission is preferentially polarized perpendicular to the plane-of-sky component of the magnetic field.  These dust grains are typically at temperatures in the range $10-100$\,K, and so their thermal emission peaks in the far-infrared (FIR) regime.  In this work, we define the FIR regime as approximately $10-300\,\mu$m: a wavelength range which is inaccessible to ground-based astronomy due to the water in the Earth's atmosphere, and which traces the peak of the dust spectral energy distribution (SED).

This paper is laid out as follows: in Section~\ref{sec:need}, we discuss the need for PRIMAger in order to observe magnetic fields at high resolution across star-forming clouds.  In Section~\ref{sec:obs} we describe our proposed observations, and in Section~\ref{sec:time} we produce an estimate of the time required for these observations.  Section~\ref{sec:example} shows some example PRIMAger observations of a nearby molecular cloud, demonstrating the orders of magnitude improvement in mapping speed, sensitivity and resolution that PRIMAger will bring.  Section~\ref{sec:interpretation} briefly describes some of the analyses that these observations would enable.  Section~\ref{sec:summary} summarizes this work.

\section{The need for PRIMAger}
\label{sec:need}

\begin{figure}
    \centering
    \includegraphics[width=1\linewidth]{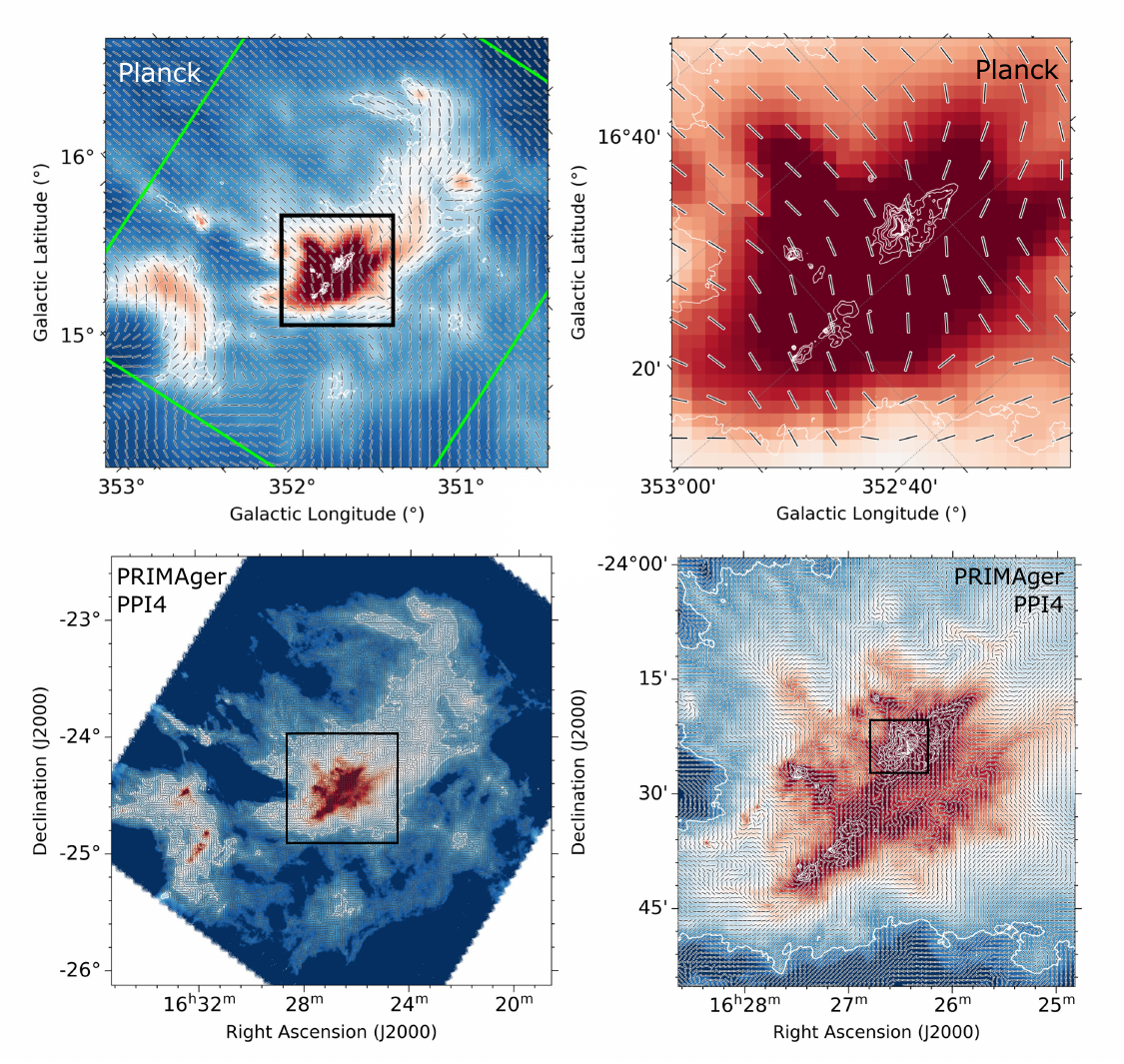}
    \caption{A comparison of \textit{Planck} Observatory 353\,GHz (850\,$\mu$m) dust polarization observations of the Ophiuchus molecular cloud\cite{planck2015} with model PRIMAger Band PPI4 (215\,$\mu$m) observations, derived from \textit{Herschel} Gould Belt Survey 250$\mu$m data\cite{ladjelate2020}.  These model observations demonstrate the factor 11--28 (PPI4--PPI1) improvement in angular resolution that PRIMAger will bring to the mapping of magnetic fields across molecular clouds, and the area over which we would expect to achieve a 5-$\sigma$ detection in polarized light.  \textit{Upper left:} \textit{Planck} plane-of sky magnetic field vectors, plotted every 5$^{\prime}$ over the \textit{Planck} dust emission map.  The green rectangle marks the footprint of the \textit{Herschel} observations shown in the lower left panel. The black rectangle marks the zoomed in region shown in the upper right panel. \textit{Herschel}/SPIRE 250\,$\mu$m contours are overlaid in white at levels of 83, 1200, 1500, 2000, 2500, 4000, 5500 and 7000 MJy/sr. \textit{Upper right:} The intermediate-mass L1688 star-forming region, as observed by \textit{Planck}; contours are as in the left-hand panel. \textit{Lower left:} Model PRIMAger Band PPI4 magnetic field vectors, plotted over the \textit{Herschel} 250\,$\mu$m map\cite{ladjelate2020}.  Vectors are plotted where we expect to achieve a 5-$\sigma$ detection in polarized light. The black rectangle and the contours match those in the upper left panel. \textit{Lower right:} Model PRIMAger Band PPI4 vectors in the L1688 region. Model magnetic field vectors are plotted every 36$^{\prime\prime}$ for clarity, comparable to the 27.6$^{\prime\prime}$ resolution of the PPI4 band.  The area plotted in Figure~\ref{fig:fig2} is marked with a black box.  Contours are as in the upper left panel.}
    \label{fig:fig1}
\end{figure}

There has never yet been a space-based telescope sensitive to far-infrared polarization.  The \textit{Planck} Observatory observed polarized dust emission in the submillimetre regime, which we define as approximately $300-1000\,\mu$m, tracing the Rayleigh-Jeans tail of the dust SED. \textit{Planck} produced maps of linearly polarized Galactic dust emission at 850$\mu$m (353\,GHz) \cite{planck2015}, with 5$^{\prime}$ resolution.  This is sufficient to measure the mean magnetic field direction across nearby molecular clouds, but not to resolve the magnetic fields within star-forming filaments and cores, as shown in the upper panels of Figure~\ref{fig:fig1}.

A few bands in the submillimetre are observable from the ground at the best observing sites on Earth.  At these locations, large single-dish observatories can observe dust emission at resolutions of approximately 10$^{\prime\prime}$.
Notably, the POL-2 polarimeter\cite{friberg2016} on the James Clerk Maxwell Telescope (JCMT) has observed high-column-density regions within molecular clouds in polarized light at 850$\mu$m with 14$^{\prime\prime}$ resolution.  Many of these observations have been made under the auspices of the JCMT BISTRO (B-fields in Star-forming Region Observations) Survey\cite{wardthompson2017}, which has observed a wide range of nearby dense cores\cite{liu2019,coude2019,eswaraiah2021,karoly2023}, low- and high-mass filaments\cite{pattle2017,doi2020,kwon2022}, and higher-mass hub-filament systems\cite{arzoumanian2021,hwang2022,ching2022a} at distances $\lesssim 2$\,kpc, with an extension to distances $\lesssim 8$\,kpc (the Galactic Centre) ongoing\cite{karoly2025,yang2025}.  However, the total area mapped by the BISTRO survey is only approximately $1.5$\,deg$^{2}$, and POL-2 observations are limited to regions of high surface brightness by atmospheric constraints.  Higher resolution dust polarization observations can be achieved with submillimetre interferometers, such as the Submillimeter Array (SMA)\cite{tang2009,zhang2014,ching2017} and the Atacama Millimeter/submillimeter Array (ALMA) \cite{sadavoy2018,fernandezlopez2021,sanhueza2021}.  However, the small field of view and loss of large-scale structure inherent to interferometers prevents their use as large-scale survey instruments for the interstellar medium.

In contrast, stratospheric observatories can successfully observe at FIR wavelengths. The most common stratospheric platforms are balloons, of which one of the most recent iterations is the BLASTPol\cite{fissel2010} telescope.  BLASTPol observed in polarized light at 250$\mu$m, 350$\mu$m and 500$\mu$m, but was restricted to observing a limited number of very low-declination clouds visible only from the Southern Hemisphere\cite{gandilo2016,fissel2019} due to its launch site in Antarctica.

The launch site restrictions faced by balloon-borne observatories can be avoided by placing telescopes on powered aircraft.  The airborne Stratospheric Observatory for Infrared Astronomy (SOFIA)\cite{reinacher2018}, a modified Boeing 747 aircraft carrying a 2.5\,m mirror, could observe star-forming regions in the FIR from both hemispheres. SOFIA's High-resolution Airborne Wideband Camera Plus (HAWC+)\cite{harper2018} instrument provided the highest-resolution and shortest-wavelength far-infrared polarization observations to date, operating at wavelengths 50--240\,$\mu$m, with resolutions comparable to those that PRIMAger will achieve.  Before SOFIA was decommissioned in 2022, HAWC+ was used to observe a range of star-forming clouds at FIR wavelengths\cite{chuss2019,santos2019,pillai2020,stephens2022}.  

While it could access FIR wavelengths, HAWC+ nevertheless suffered from similar limitations to ground-based submillimetre instruments due to atmospheric constraints, which limited its sensitivity, mapping speed, and response to large-scale structure.  The \textit{Herschel} Space Observatory, which operated at wavelengths 70--500\,$\mu$m, demonstrated the step change in far-infrared imaging that can be achieved by a space telescope compared to ground-based or stratospheric observatories.  \textit{Herschel} mapped entire molecular clouds down to the cirrus confusion limit\cite{andre2010}: these multiwavelength observations were used to characterize the dust SED, thereby creating maps of column density and dust temperature across molecular clouds\cite{konyves2015,ladjelate2020}, and to perform statistical analyses of their properties and comparisons between clouds\cite{schneider2013}.  Mapping to these depths across the same areas in polarized light with PRIMAger would allow us to perform the same statistical analyses on the magnetic fields of molecular clouds.

\begin{figure}
    \centering
    \includegraphics[width=1\linewidth]{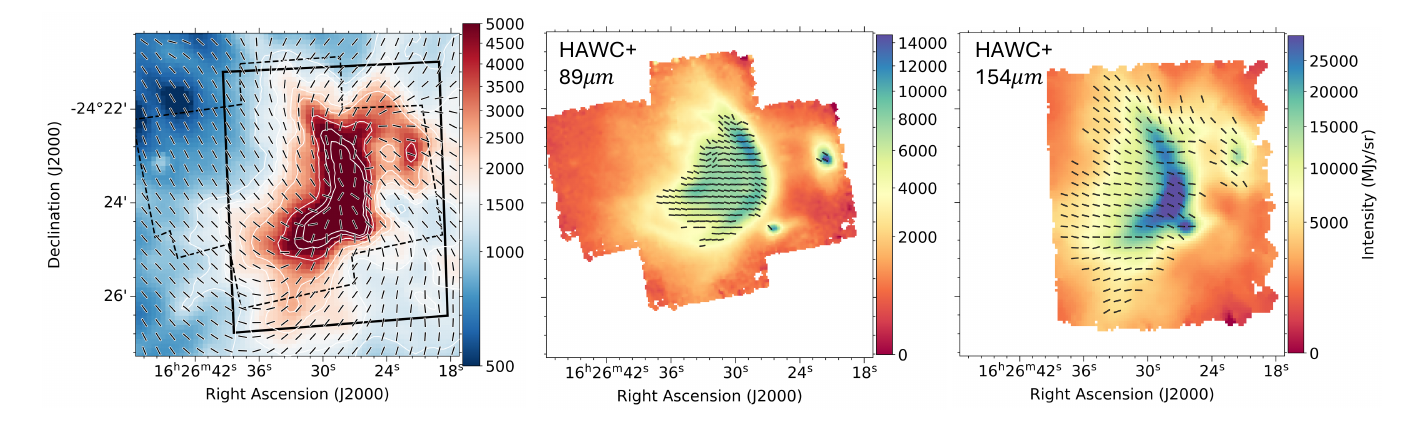}
    \caption{A comparison of our model PRIMAger observations with the current best available far-infrared polarization observations of the Oph A region, made with the HAWC+ instrument on SOFIA\cite{santos2019}.  (Comparable-resolution observations over a slightly larger area have been made at 850$\mu$m using POL-2\cite{kwon2018}.)  Each panel shows the same area of the sky. \textit{Left:} Our model PRIMAger Band PPI4 observations for Oph A. The HAWC+ 89\,$\mu$m observing footprint is marked with a black dashed line. The HAWC+ 154\,$\mu$m observing footprint is marked with a solid black line. Contours are as in Figure~\ref{fig:fig1}.  Vectors are plotted with 18$^{\prime\prime}$ spacing, a resolution intermediate between PRIMAger Bands PPI2 and PPI3. \textit{Center:} The HAWC+ 89\,$\mu$m Stokes $I$ map, with magnetic field vectors overplotted. Vectors are plotted with 8$^{\prime\prime}$ spacing, to reflect the HAWC+ 89\,$\mu$m resolution, and have SNRs of $I/\sigma_{I} \geq 200$ and $p/\sigma_{p} \geq 3$. \textit{Right:} The HAWC+ 154\,$\mu$m Stokes $I$ map, with magnetic fields vectors overplotted with 14$^{\prime\prime}$ spacing to represent the HAWC+ 154\,$\mu$m resolution; SNR cuts are as in the central panel.}
    \label{fig:fig2}
\end{figure}

Figures~\ref{fig:fig1} and \ref{fig:fig2} show the improvement that PRIMAger would bring over previous space-based and stratospheric polarization mapping.  As shown in Section~\ref{sec:example}, PRIMAger would result in a $>10\times$ improvement in angular resolution over \textit{Planck}, and a $\sim 10\,000\times$ increase in mapping speed over HAWC+.  Unlike \textit{Planck}, PRIMAger will measure the peak of the dust SED in multiple bands, thereby probing dust composition and disentangling line of sight complexity.  PRIMAger will thus allow us to for the first time produce unbiased surveys of magnetic fields within entire molecular clouds.

\section{Proposed observations}
\label{sec:obs}

We propose to systematically map all of the star-forming clouds within 0.5 kpc in polarized light, in order to have a complete sample of magnetic field morphologies in star-forming regions, observed from cloud to core scale.  Our preferred mapping strategy would involve making wide-area maps, using tiles of $\sim$1 square degree or greater.

We have based our proposed observations on the areas targeted by the \textit{Herschel} Gould Belt Survey\cite{andre2010}, which mapped nearby molecular clouds down to the cirrus confusion limit.  The \textit{Herschel} Gould Belt Survey mapped an area of approximately 160 deg$^{2}$\cite{andre2010}; for comparison, the largest extant polarimetric survey of molecular clouds at sub-arcminute resolution is the JCMT BISTRO Survey\cite{wardthompson2017}, which has mapped a total area of approximately $1.5$\,deg$^{2}$ in 672 hours.  Our proposed observations would thus provide a factor of approximately 100 increase in mapping area over the current state of the art in the submillimetre, and would provide multi-wavelength observations spanning the peak of the dust emission SED. 

Our observations would have a linear resolution of 0.006 pc in Band PPI1 and 0.016 pc in Band PPI4 in the nearest clouds, at a distance of approximately 130 pc: Ophiuchus, Taurus, Corona Australis and the Pipe Nebula\cite{zucker2019}.  We would achieve a linear resolution of 0.024 pc -- 0.060 pc at a distance of 500 pc.  Thus, even in the higher-mass and more distant clouds in our sample, and particularly in the high-mass Orion A and B clouds at distances of around $400$\,pc\cite{zucker2019}, we would be able to resolve the magnetic field within individual star-forming filaments and cores, which have typical size scales $\sim$0.1 pc\cite{pineda2023}.

Although our sample would include some high-mass star-forming regions (Orion A and B, and Serpens Aquila), we note that there is a lack of infrared-dark clouds (IRDCs) and other extremely massive star-forming regions within 0.5 kpc.  This could easily be addressed by scaling this project up to also map more distant targets.  In this case, targets could be selected from the Herschel HOBYS\cite{motte2010} and/or HiGAL\cite{molinari2016} surveys, or from surveys of large-scale Galactic filaments\cite{zucker2018}. 

\section{Time estimate}
\label{sec:time}

We base our estimates on \textit{Herschel} Space Observatory observations of nearby molecular clouds.  \textit{Herschel} Gould Belt Survey\cite{andre2010} SPIRE (Spectral and Photometric Imaging Receiver) maps are confusion-limited in nearby star-forming regions.  The cirrus confusion level is $\sim$30 mJy/18$^{\prime\prime}$ beam = 5 MJy/sr at 250\,$\mu$m\cite{andre2010}.  In the following estimates we take Herschel SPIRE 250\,$\mu$m brightnesses to be comparable to PRIMAger PPI4 (235\,$\mu$m) Stokes $I$ values.  We perform our sensitivity calculations for the most stringent case, in which we wish to detect polarized emission down to the cirrus confusion limit.

Polarized intensity $P$ is given by
\begin{equation}
P = \sqrt{Q^2 + U^2},
\end{equation}
where $Q$ and $U$ are the Stokes parameters.  In practice, the directly measured value of $P$ must be debiased in order to correct for non-Gaussian noise effects \cite{wardle1974,montier2015}.  The polarization fraction $p$ is then given by
\begin{equation}
    p = P/I,
\end{equation}
where $I$ is the total intensity, also referred to in this context as the Stokes $I$ parameter.

For completeness we note that polarization angle is given by
\begin{equation}
\theta = \frac{1}{2}\arctan\left(\frac{U}{Q}\right),
\end{equation}
using the IAU convention for the sign on Stokes $U$\cite{iau}, and plane-of-sky magnetic field direction is thus $\theta + 90^{\circ}$.

The maximum polarization fraction in the diffuse ISM is $\sim$20\%\cite{planck2015}.  However, polarization fractions typically decrease with increasing visual extinction, and at the highest gas densities can be as low as 2--5\%\cite{pattle2019}.  Assuming 20\% polarization in the most diffuse regions of molecular clouds, we expect that the cirrus confusion limit in total intensity (5\,MJy/sr) will have a surface brightness in polarized intensity of 1 MJy/sr.  Our target is thus to achieve a good detection in polarized intensity where $P=1$\,MJy/sr.  This would also ensure a detection of polarized emission at all higher Stokes $I$ surface brightnesses, as $p = (P/I) \propto I^{-\alpha}$, where $0 < \alpha < 1$\cite{whittet2008}.

We used the PRIMAger Exposure Time Calculator (ETC)\footnote{\url{https://prima.ipac.caltech.edu/page/etc-calc}} to calculate the telescope time required for a detection of 20\% polarized cirrus emission with 5-$\sigma$ sensitvity in PRIMAger Band PPI4.  To do so, we took 1 MJy/sr = 14.1 mJy/beam for a PPI4 beam FWHM of 27.6$^{\prime\prime}$.

The \textit{Herschel} Gould Belt Survey covered an area of $\sim$160 deg$^{2}$\cite{andre2010}.  To map this area in polarized light to the confusion limit would thus require 170 hours of telescope time for 5-$\sigma$ sensitivity to 20\% polarized cirrus in PPI4.  We summarize our calculations, and the sensitivities which would be achieved in bands PPI1--PPI4, in Table~\ref{tab:sensitivity}.  Our time estimate demonstrates that PRIMAger would provide a factor $\sim10,000$ increase in mapping speed over HAWC+ in the FIR (see discussion in Sec.~\ref{sec:example}), while also providing unprecedented sensitivity to low-surface-brightness, extended polarized emission from molecular clouds.

We note that this time estimate could easily be scaled up or down either in terms of number of clouds to be mapped, area to be mapped within molecular clouds, or desired polarization sensitivity.

\section{Model PRIMAger observations of magnetic fields in a nearby molecular cloud}
\label{sec:example}

Figure~\ref{fig:fig1} shows a model PRIMAger observation of the Ophiuchus molecular cloud, in comparison to the current best available mapping of magnetic fields on molecular cloud scales, made with the \textit{Planck} Observatory.  The upper panels of Figure~\ref{fig:fig1} show the dust emission and plane-of-sky magnetic field directions derived from \textit{Planck} observations\footnote{Data retrieved from \url{https://pla.esac.esa.int/}.}, at 850\,$\mu$m (353\,GHz).  The magnetic field segments are shown at their native 5$^{\prime}$ resolution.  The lower panels demonstrate the improvement in resolution that PRIMAger will provide.  To do so, we use \textit{Herschel} 250\,$\mu$m observations\cite{ladjelate2020} to produce a hypothetical magnetic field map for the Ophiuchus molecular cloud.  This wavelength was chosen as a reasonable proxy for the PRIMAger PPI4 band.  We emphasize that this is not a prediction of the magnetic field morphology in the Ophiuchus molecular cloud, and is simply a plausible magnetic field geometry intended to demonstrate the capabilities of PRIMAger.

We created a model magnetic field by assuming that magnetic fields transition from being preferentially parallel to density structure (i.e. perpendicular to the local column density gradient) at low column densities to being preferentially perpendicular to density structure (i.e. parallel to the local column density gradient) at high column densities\cite{soler2013,planck2015,soler2019}.  We assumed that 250$\mu$m intensity is proportional to column density (note that this is strictly true only for isothermal emission), and so constructed artificial Stokes $Q$ and $U$ maps from the 250$\mu$m Stokes $I$ map.  We arbitrarily chose to centre the parallel-to-perpendicular transition on a 250$\mu$m intensity of 950\,MJy\,sr$^{-1}$, and used a sigmoid smoothing function to prevent a sharp jump in magnetic field angles.  The model field geometry is shown in the lower panels of Figure~\ref{fig:fig1}.  In the lower right-hand panel, vectors are sampled every 36$^{\prime\prime}$ to approximate the resolution of the PRIMAger PPI4 band.  Vectors are plotted at Stokes $I$ brightnesses of 1 MJy/sr and above (the surface brightness above which we aim to achieve a detection in polarized light; cf. Section~\ref{sec:time}).  It can immediately be seen that PRIMAger will produce more than an order of magnitude improvement in angular resolution over \textit{Planck}, and has the sensitivity to detect magnetic fields into the cirrus emission at the peripheries of molecular clouds.

In Figure~\ref{fig:fig2}, we compare our model PRIMAger observations to HAWC+ observations\cite{santos2019} at the same location, to demonstrate the orders of magnitude increase in mapping speed that PRIMAger will bring to far-infrared polarimetry. The 89$\mu$m HAWC+ oberservations took approximately 4 hours, while the 154$\mu$m observations took approximately 2 hours\cite{santos2019} to achieve mean Stokes $I$ RMS sensitivities of 215 MJy/sr at 89$\mu$m and 43 MJy/sr at 154$\mu$m\cite{le2024} (sensitivities in $P$ are not given in these works).  The area mapped by HAWC+ at each wavelength is very approximately 25 sq. arcmin per map.  The PRIMA ETC suggests that PRIMA could achieve a 5-sigma sensitivity of 0.11 MJy/sr over 25 sq. arcmin in 4 h, i.e. a factor $\sim$10,000 increase in mapping speed over the current state of the art for far-infrared polarizarion observations.

\begin{table}[]
    \centering
    \begin{tabular}{l cccc}
        \hline
        \hline
        PRIMAger Filter & PPI1 & PPI2 & PPI3 & PPI4 \\
        \hline
        Wavelength ($\mu$m) & 96 & 126 & 172 & 235 \\ 
        Cirrus confusion level (MJy/sr) & -- & -- & -- & 5.0 \\
        Assumed \% pol. in cirrus & -- & -- & -- & 0.2 \\
        Polarized intensity in cirrus (MJy/sr) & -- & -- & -- & 1.0 \\
        \hline
        Time required for 5-$\sigma$ detection in 1 deg$^{2}$ tile (h) & -- & -- & -- & 1.1 \\
        Time required for 5-$\sigma$ detection over 160 deg$^{2}$ (h) & -- & -- & -- & 170 \\
        \hline
        5-$\sigma$ sensitivity achieved in $P$ (MJy/sr) & 2.52 & 1.94 & 1.37 & 1.00 \\
        \hline
        \hline
    \end{tabular}
    \caption{This table lists our estimated time (including overheads) required to achieve a 5-$\sigma$ detection of 20\% polarized cirrus emission over a 160 deg$^2$ area in PRIMAger Band PPI4, and the 5-$\sigma$ sensitivities achieved in Bands PPI1--PPI3 in the same amount of time.}
    \label{tab:sensitivity}
\end{table}

\section{Interpretation}
\label{sec:interpretation}

In this section, we briefly describe some of the analyses that we would perform on the wide-area, multi-wavelength maps of plane-of-sky magnetic field morphology that our survey would produce.  While we detail some specific analyses below, our overarching aim is to determine whether magnetic field properties change how star formation proceeds in a molecular cloud, and particularly whether it changes the star formation rate or efficiency of a cloud, or how the cloud fragments into gravitationally unstable star-forming clumps and cores.  We will compare the observed magnetic fields to the results of numerical simulations of both weak- and strong-field star formation, which will have undergone radiative transfer post-processing\cite{reissl2016} to produce results which are directly comparable to our observations.  

These polarization observations would also give us exquisitely deep maps of all of the clouds in our survey in total intensity, which would allow us to pursue a range of science goals which are beyond the scope of this paper, such as performing a census of candidate pre-brown dwarf cores in nearby star-forming regions clouds\cite{palau2024}, or exploring the properties of extragalactic background sources\cite{difrancesco2020}.

\subsection{Deriving magnetic field properties}

In order to quantitatively determine the importance of magnetic fields in the interstellar medium, their strength must be measured.  Magnetic field strength values are required to calculate all of the key diagnostics of the dynamic importance of magnetic fields: the mass-to-flux ratio, which assesses the relative importance of magnetic fields and gravity; the Alfv\'en Mach Number, comparing magnetic fields and non-thermal motions; and plasma $\beta$, comparing magnetic fields and thermal pressure\cite{pattle2023}.  

While dust polarization observations provide only a plane-of-sky magnetic field direction, these measurements can be used to estimate plane-of-sky magnetic field strength in molecular clouds using a modern implementation of the Davis-Chandrasekhar-Fermi (DCF) method\cite{davis1951,chandrasekhar1953}. The DCF method uses dispersion in polarization angle as a proxy for Alfv\'enic Mach number, and so, for a measured gas velocity dispersion and density, allows inference of POS magnetic field strength.  Recent modifications of the DCF method include use of structure function methods to characterize angle dispersion\cite{hildebrand2009,houde2009}, or both angle and velocity dispersion\cite{lazarian2022}, thereby allowing a wider range of spatial scales over which to calculate magnetic field statistics.  Another more radical reformulation proposes taking non-Alfv\'enic (compressible)
turbulent modes to dominate over Alfv\'enic (incompressible) modes\cite{skalidis2021}, although this approach remains under debate in the literature\cite{liu2021,li2022}.  Our large mapping area will give us an unprecedentedly large ensemble of polarization measurements, which will allow us to investigate the efficacy of the plethora of DCF variants\cite{hildebrand2009,houde2009,pillai2015,cho2016,pattle2017,hwang2021,skalidis2021,liu2021,lazarian2022,li2022} in a wide range of environments. 

Our large ensemble of measurements will also allow us to marginalize over projection effects and so to statistically investigate the total (rather than plane-of-sky only) magnetic strengths in molecular clouds\cite{crutcher2004}.  We will also use alternative, polarization fraction-based, methods for deriving three-dimensional magnetic field strengths\cite{chen2019,hu2023}, and will be able to compare the results of these different approaches.  We discuss further methods for directly obtaining three-dimensional magnetic fields through synthesis with forthcoming radio-wavelength observations in Section~\ref{sec:synergies}, below.  DCF estimations of magnetic field strength can further be compared to field strength estimates made using the `intensity gradient' approach, which compares the magnetic field direction to the dust emission gradient in regions of significant self-gravity\cite{koch2012,koch2014}.

A critical aspect of DCF analysis is choosing an appropriate line tracer with which to accurately characterize the velocity dispersion in the dense gas that the dust polarization emission is tracing\cite{fissel2019}.  Our observations will cover a wide range of gas densities and star-forming environments, and so it is vital to have dense gas tracers with a wide range of critical densities\cite{difrancesco2007} available to us in order to accurately determine magnetic field strengths across entire molecular clouds.  Moreover, DCF implicitly requires that POS magnetic fields are estimated over an area comparable to the LOS depth of the cloud being sampled by the line tracer used\cite{ostriker2001}.  The high sensitivity of PRIMA, and its higher resolution compared to previous space-based polarimetric measurements made with \textit{Planck}, will allow magnetic field fluctuations to be measured over a wider range of angular scales than has previously been possible.  Although the requirement for LOS depth and POS scale to match can be mitigated against by use of structure-function-based DCF implementations\cite{lazarian2022}, having a wide variety of line tracers available to us, with a range of critical densities, is crucial for our analysis.

Ground-based millimetre and submillimetre-wavelength telescopes such as the JCMT, Green Bank, APEX and the IRAM 30m telescope are currently providing us with wide-area mapping of line tracers covering a broad range of critical densities with angular resolutions comparable to those that PRIMA will achieve, including various rotational transitions of CO and its isotopologues\cite{rigby2016,schuller2017,goicoechea2020,park2023}, NH$_3$\cite{friesen2017}, and a wide range of other species, including but not limited to HCO$^+$, HCN, CS, SO and N$_2$H$^+$\cite{fuente2019}.  In the 2030s, the proposed Atacama Large Aperture Submillimeter Telescope (AtLAST) will map the chemistry and dynamics of star-forming regions with unprecedented sensitivity and resolution\cite{klaassen2024}, providing a further resource with which to calibrate our DCF measurements across star-forming environments.

It will thus be possible to use our dust polarization observations to determine the energy balance between gravity, magnetic fields, non-thermal gas motions and stellar feedback in the molecular clouds that we observe.  By observing magnetic fields at high resolution across entire clouds, we will be able to systematically investigate how the dynamic importance of magnetic fields changes as a function of temperature, density and radiation environment, as well as other environmental factors.

\subsection{Magnetised cloud formation models}

How molecular clouds form, and the role that magnetic fields play in that process, remain poorly constrained by observations.  Our observations can be used to test molecular cloud formation mechanisms by comparing our observed magnetic field morphologies and strengths to radiative transfer post-processed simulations of molecular cloud formation.  Some cloud formation models, particularly cloud-cloud collision models\cite{inutsuka2015}, already provide testable predictions of magnetic field geometry.  By the time that PRIMA is operational, there will also be clear predictions from other models, such as models in which molecular clouds form from large-scale turbulence\cite{walch2015,kim2017,hennebelle2018}, and Parker Instability\cite{parker1966,rodrigues2016} models.

\subsection{The relationship between magnetic fields and density structure}

The relationship between magnetic fields and density structure in molecular clouds can be quantified using the Histogram of Relative Orientations (HRO)\cite{soler2013}, and other similar analyses\cite{jow2018}.  These techniques have also been extended to assess the relationship between magnetic fields and filamentary structure, and H\textsc{ii} region structure\cite{khan2024}, and to assess the magnetic field geometry in individual star-forming cores\cite{perry2025}.  More detailed models exist for the collapse of star-forming cores proceeds in the presence, or absence, of a strong magnetic field\cite{myers2021}.

Magnetic fields are expected to be preferentially aligned parallel to gas structures at low column density, and perpendicular at high column density\cite{soler2019}.  This transition is likely related to the onset of gravitational instability in the cloud\cite{soler2017,seifried2020}.   Our proposed observations would allow us to test this hypothesis by investigating if and how the critical column density of this transition varies with molecular cloud properties, magnetic field strengths and energetic balance.

Observations of magnetic fields within filaments\cite{doi2020} and cores\cite{myers2021} have been compared to plane-of-sky projections of three-dimensional model fields.  However, small sample sizes currently prevent a good statistical understanding of the three-dimensional magnetic field structures that are consistent with our plane-of-sky observations.  Each of the larger clouds in the Gould Belt contains hundreds of filamentary structures and dense cores\cite{konyves2015}, and so our high-resolution observations of magnetic fields over entire clouds would also allow us to systematically investigate how these fields behave in a wide variety of ISM structures, with significant statistical power.  Our multi-wavelength imaging will also allow us to trace dust populations at different temperatures, and so to investigate how magnetic fields vary as a function of density, as we describe in Section~\ref{sec:mwl}.

\subsection{The relationship between magnetic fields and star formation feedback}

We will investigate proposed relationships between magnetic fields, (proto)stellar feedback and star formation efficiency\cite{krumholz2019} through comparison with star formation rates measured from mid-IR source number counts, and with gas dynamics inferred from spectroscopic observations.

Interferometric observations suggest that there is typically little or no correlation between magnetic field and outflow directions on $\sim 1000$\,au scales\cite{hull2013,hull2019}, although there is a subset of sources with good alignment\cite{huang2024}.  However, observations made with JCMT POL-2 and with the \textit{Planck} Observatory suggest that magnetic fields may be systematically misaligned with respect to outflows in three-dimensional space on $\sim$0.1\,pc scales\cite{yen2021,xu2022}.  In at least some cases, outflows show a preferentially perpendicular orientation with respect to filamentary structure, which suggests an important role for magnetic fields in setting their directions\cite{stephens2017,kong2019}.  Our proposed observations would allow us to perform a full census of magnetic fields in the vicinity of protostellar outflows in nearby molecular clouds.

Our multi-wavelength polarization observations will also allow us to investigate magnetic fields under intense stellar feedback in photodissociation regions (PDRs), as we describe in Section~\ref{sec:mwl}.

\subsection{Testing of alternative methods for inferring ISM magnetic fields}

Our observations will allow testing of alternative (non-polarimetric) methods for determining the properties of magnetic fields in the ISM.

Observations of H\textsc{i} emission in the low-density ISM have shown that thin structures in velocity channel maps preferentially align parallel to the magnetic field \cite{clark2014,clark2015}.
While debate continues about the origin of this effect\cite{clark2019,kalberla2020,yuen2021,hu2023a,kalberla2024,yuen2024}, the Velocity Gradient Technique (VGT; \cite{gonzalez2017,yuen2017}) has used this correlation to infer ISM magnetic field directions from unpolarized spectral line measurements in the CNM and in molecular clouds.  VGT has been used to reconstruct magnetic fields on the large scales observed by \textit{Planck}\cite{hu2019,schmaltz2023,schmaltz2024}, and modifications accounting for self-gravity\cite{yuen2017b} or feedback\cite{liu2024a} have been presented.  However, a lack of correlation between VGT predictions and dust polarimetry measurements has been seen in some cases\cite{hsieh2019,soam2024}.

Alternative paradigms use the ``magnetic coherence'' of H\textsc{i} filaments identified in velocity channel maps to investigate magnetic field properties.  One such method uses H\textsc{i} velocity cubes to reconstruct three-dimensional magnetic fields in the CNM\cite{clark2019a}, using the coherence of H\textsc{i} velocity structures as a measure of the degree of LOS magnetic field tangling\cite{clark2018}.  However, in star-forming regions within molecular clouds, little correlation is seen between velocity gradients and magnetic field directions\cite{chen2024}.

An advantage of these spectral-line-based techniques is that where they are applicable, they can probe the magnetic field structures of multiple velocity components along a single LOS\cite{clark2018,hu2019}.  Our high resolution dust polarization observations across entire molecular clouds from low and high densities will allow robust and systematic investigation of the regime(s) over which these techniques can be used to map magnetic field directions in molecular clouds using unpolarized spectral line data.  Where alignment between magnetic fields and velocity channel structure breaks down may further serve as a probe of the transition of the ISM to the gravity-dominated regime\cite{hu2020,hu2021}, although gradient orientations can also be altered by shocks\cite{chen2024}.

Our observations will further provide tests and points of comparison for machine learning-based methods for determining magnetic field structure from velocity data\cite{xu2023,hu2024}, as well as serving as input into machine learning models that incorporate polarization measurements into their inference of ISM magnetic field properties\cite{xu2025}.

\subsection{Dust properties from polarization measurements}
\label{sec:mwl}

\subsubsection{Grain alignment in molecular clouds}

PRIMAger will play a pivotal role in our understanding of the alignment of dust grains in molecular clouds.  Over the vast majority of the ISM, dust grains are aligned with respect to the magnetic field\cite{planck2020,hoang2022,tram2022}.  This alignment is understood to arise at least in part from Radiative Alignment Torques (RAT)\cite{lazarian2007}; in this paradigm, dust grains gain angular momentum from anisotropic incident radiation fields, resulting in paramagnetic grains precessing around the local magnetic field direction.  However, a corollary of this theory is that as $A_V$ increases, a loss of efficiency of grain alignment with respect to the magnetic field can be expected due to the decreased interstellar radiation field (ISRF)\cite{andersson2015}.  This is consistent with the widely observed `polarization hole' effect in which polarization fraction is observed to decrease with increasing $A_V$ in molecular clouds\cite{whittet2008,alves2014,jones2015}.  Nevertheless, some degree of grain alignment persists even to the highest densities in starless cores, with the degree of alignment retained appearing to be related to the strength of the local ISRF \cite{pattle2019}.  This means that the observed magnetic field morphology from dust emission polarimetry is an integration of the polarized emission from dust grains distributed along the line of sight (LOS), weighted by the grain alignment efficiency along the LOS.  Characterization of this weighting is thus crucial for accurate inference of POS magnetic field properties in molecular clouds.  High-resolution single-dish observations are already being used to investigate how grain alignment varies across molecular clouds\cite{ngoc2023,ngoc2024,pravash2025}, but are restricted to relatively small, high-surface-brightness areas.  Our high-resolution, wide-area, multiwavelength observations will therefore allow us to for the first time investigate how well grains are aligned with respect to the magnetic field as a function of $A_V$ and ISRF across entire molecular clouds.

In very high-density and/or very highly-irradiated environments, alternative grain alignment mechanisms may come into play, including precession of grains around the radiation vector\cite{tazaki2017}, or the electric vector\cite{lazarian2020}, or alignment by mechanical torques\cite{hoang2018}.  The extreme environments in which these mechanisms are generally observable only with interferometers, such as in protostellar discs\cite{hull2022}, or in the immediate vicinities of massive protostars\cite{pattle2021}, although some hints of precession around the electric vector have been seen in single-dish observations of the envelope of an AGB star\cite{andersson2024}. Nonetheless, we will be able to search for signs of alternative grain alignment mechanisms in our observations, particularly in high-mass star-forming regions such as Orion A and B and Serpens W40.

Particularly strong radiation fields may spin grains up sufficiently that their tensile strength can no longer hold them together\cite{silsbee2016}.  This effect, known as Radiative Torque Disruption (RAT-D), is predicted to lead to depolarization in highly irradiated regions where a strong polarization signal would otherwise be expected\cite{hoang2021}.  Hints of RAT-D have been seen in previous single-dish polarization observations, particularly in the Orion Molecular Cloud\cite{legouellec2023,ngoc2024,tram2024}, although some recent work suggests their efficacy may be lower than previously predicted\cite{silsbee2025}.  Our proposed observations will therefore allow us to investigate the environments in which RAT-D is predicted to become significant, with PRIMAger's multiwavelength observations providing us with a powerful tool for determining the strength of this effect and disentangling which dust grain populations are most affected\cite{tram2024}.

\subsubsection{Dust composition from multiwavelength observations}

A key advantage of PRIMAger will be simultaneous observations at multiple FIR wavelengths in polarized light, spanning the peak of the dust emission SED.  This will allow us to produce polarization spectra -- measurements of polarization percentage as a function of wavelength -- across entire molecular clouds, against which dust composition models can be tested\cite{lee2024,tram2024}.

There is significant variation in polarization spectrum shape both between and within clouds.  Some previous submillimetre and FIR observations have suggested a minimum in polarization fraction at 350\,$\mu$m\cite{vaillancourt2008}, while elsewhere a flat polarization spectrum has been seen\cite{gandilo2016,ashton2018}.  HAWC+ observations have shown that far-infrared polarization spectra are variable within molecular clouds\cite{santos2019,michail2021}, and find correlations between the slope of the polarization spectrum and both temperature and density within clouds.  These observations provide important tests for dust composition models\cite{guillet2018,hensley2023}, as well as for the Radiative Alignment Torques model of grain alignment\cite{lee2024,tram2024}.  By observing polarization spectra over entire clouds, we will be able to test models of dust composition in the diffuse peripheries of clouds, and evaluate how dust properties evolve from low to high densities in molecular clouds.

In regions with particularly strong temperature gradients, different dust populations are traced in different FIR bands, with shorter wavelengths tracing a hotter dust population\cite{legouellec2023}.  This is apparent in Figure~\ref{fig:fig2}, in which it can be seen that HAWC+ 89$\mu$m observations preferentially trace the hot dust associated with the star S1, while the 154\,$\mu$m observations preferentially trace the bulk of the dust mass in the dense Oph A clump.  Systematically observing nearby high-mass star-forming regions in polarized light -- particularly Orion A and B, and Serpens Aquila -- will give us important insights into the role of magnetic fields in the evolution of PDRs\cite{hwang2023}, including in synergy with observations of the [C\textsc{ii}] 158\,$\mu$m and [O\textsc{i}] 63\,$\mu$m fine structure cooling lines\cite{pabst2017,karim2023}.   An extension of our proposed survey to higher-mass star-forming regions within 2\,kpc would allow magnetic fields in a larger sample of HII regions to be mapped\cite{pattle2018,tahani2023}.

\subsection{Synergies with other facilities to construct three-dimensional magnetic field morphologies}
\label{sec:synergies}

As we have discussed, PRIMA will provide a unique and unparalleled opportunity to study interstellar magnetic fields in the FIR regime.  However, it will also have a range of synergies with ground-based instrumentation, particularly the forthcoming Square Kilometre Array (SKA)\cite{dewdney2009} and the proposed next-generation Very Large Array (ngVLA)\cite{murphy2018}.  These instruments will map non-thermal sources of polarization, notably synchrotron emission\cite{dickinson2015}, and will observe line-of-sight magnetic fields in the diffuse ISM and the peripheries of molecular clouds using Faraday rotation techniques\cite{tahani2018}.  The SKA and the ngVLA, and the forthcoming ALMA Wideband Sensitivity Upgrade (WSU)\cite{carpenter2023}, will also provide unprecedented sensitivity to Zeeman splitting of the spectral lines of paramagnetic species, giving direct access to line-of-sight magnetic field strengths\cite{crutcher2019,robishaw2015}.  These instruments will therefore have the ability to observe line-of-sight magnetic field strengths in a range of environments, complementing our plane-of-sky mapping.  Combining SKA or ngVLA Zeeman and Faraday rotation observations with our PRIMAger plane-of-sky magnetic field mapping will provide an unprecedented opportunity to construct three-dimensional magnetic field morphologies in star-forming molecular clouds.

\section{Summary}
\label{sec:summary}
 
In this paper we have presented the case for an unbiased survey of star-forming molecular clouds within 0.5 kpc of the Earth in polarized light with the PRIMAger Polarimetry Imager.  This survey would map magnetic fields over entire molecular clouds at linear resolutions of $\sim10^{-3}-10^{-2}$ pc ($\sim10^{3}-10^{4}$ au) in PRIMAger Bands PPI1 -- PPI4, thereby resolving magnetic field structure both within individual star-forming filaments and cores, and in the most diffuse regions of molecular clouds. 

We estimate that the time required to map a 160 deg$^2$ area (the area observed by the \textit{Herschel} Gould Belt Survey) to the cirrus confusion limit in polarized light is 170 hours.  This would result in a 5-$\sigma$ detection of 20\% polarized cirrus emission, with surface brightnesses in polarized intensity of 1.0--2.4\,MJy/sr across the PRIMAger bands, and would also ensure detection of polarized emission at all higher column densities in the molecular cloud.  PRIMAger represents a factor $>10$ improvement in linear resolution over \textit{Planck}, the only previous space-based submillimetre polarimeter, and a factor $\sim 10,000$ increase in mapping speed over HAWC+, the current state of the art of polarimetric mapping in the FIR, while providing unprecedented sensitivity to polarized FIR emission from the diffuse outer regions of molecular clouds.

Our multi-wavelength observations of magnetic field direction would allow us to answer long-standing questions about the role of magnetic fields in star formation and the formation and evolution of molecular clouds, while our measurements of polarization fraction would allow detailed investigation of how interstellar dust grain properties vary with environment.  Our proposed survey is thus a demonstration of the unique science that PRIMA will make possible, and of how PRIMA will revolutionise our understanding of the role of magnetic fields in the star formation process.

\section*{Disclosures}
The authors declare that they have no conflicts of interest.

\section* {Code, Data, and Materials Availability} 
The code used to make Figure~\ref{fig:fig1} in this work is available from the corresponding author on request.  All data used are available from public archives, as detailed in the body of the paper.

\section* {Acknowledgments}
K.P. is a Royal Society University Research Fellow, supported by grant number URF\textbackslash R1\textbackslash 211322.  J.K. is supported by the Royal Society under grant number RF\textbackslash ERE\textbackslash 231132, as part of project URF\textbackslash R1\textbackslash 211322.  L.B.F.'s participation in this project was enabled by the in2scienceUK \textit{in2research} programme (\url{https://in2scienceuk.org}).  This research was carried out in part at the Jet Propulsion Laboratory, California Institute of Technology, under a contract with the National Aeronautics and Space Administration (NASA).


\bibliographystyle{spiejour}   


\vspace{2ex}\noindent\textbf{Kate Pattle} is a Royal Society University Research Fellow at University College London (UCL).

\vspace{2ex}\noindent\textbf{Janik Karoly} is a Postdoctoral Research Associate at UCL.

\vspace{2ex}\noindent\textbf{Lorna Buhil Findlay} is an undergraduate student at UCL.

\vspace{2ex}\noindent\textbf{Simon Coud\'e} is a Postdoctoral Researcher at Worcester State University and the Harvard \& Smithsonian Center for Astrophysics.

\vspace{2ex}\noindent\textbf{Brandon S. Hensley} is a scientist at the Jet Propulsion Laboratory, California Institute of Technology.

\vspace{2ex}\noindent\textbf{Paulo C. Cortes} is an Operations Astronomer at the Joint ALMA Observatory.

\vspace{2ex}\noindent\textbf{James Di Francesco} is Director of Optical Astronomy at Herzberg Astronomy and Astrophysics Research Centre and the Director of the Dominion Astrophysical Observatory.  He is also an Adjunct Professor in the Department of Physics and Astronomy of the University of Victoria.

\vspace{2ex}\noindent\textbf{Valentin J. M. Le Gouellec} is a postdoctoral fellow at the Institute of Space Sciences (ICE-CSIC), University Autonomous of Barcelona, Spain.

\vspace{2ex}\noindent\textbf{Enrique Lopez-Rodriguez} is an Associate Professor at the University of South Carolina.

\vspace{2ex}\noindent\textbf{Fabien Louvet} is a researcher at the Centre National de la Recherche Scientifique (CNRS) in the Universit\'e Grenoble Alpes.

\listoffigures
\listoftables

\end{document}